\documentclass[11pt, conference, letterpaper]{IEEEtran}
\usepackage{cite}
\usepackage{amsmath,amssymb,amsfonts}
\usepackage{graphicx}
\usepackage{textcomp}
\usepackage{xcolor}

\usepackage{mathtools}  
\usepackage{amsmath}
\usepackage{amssymb}
\usepackage{tabulary}
\usepackage{booktabs}
\newtheorem{theorem}{Theorem}

\usepackage[top=.75in, left=0.65in, bottom=1.1in, right=0.7in]{geometry}

\usepackage{caption}
\usepackage{subcaption}
\usepackage{booktabs}
\usepackage{bbm}
\usepackage{algorithmic}
\usepackage{algorithm}
\usepackage[numbers]{natbib}

\usepackage[hyphens]{url}
\usepackage[hidelinks]{hyperref}

\DeclareMathOperator*{\argmax}{arg\,max}

\DeclarePairedDelimiter\abs{\lvert}{\rvert}%
\DeclarePairedDelimiter\norm{\lVert}{\rVert}%
\newcommand\algName{\textit{B2P-Stream}}
\newcommand\nbTs{T}
\newcommand\maxU{K}
\newcommand\nbUsers{N}

\newcommand\bl{b}
\newcommand\bufferSize{s}
\newcommand\blEst{\Tilde{\bl}}
\newcommand\rEst{\Tilde{\mu}}
\newcommand\expR{\mu}

\newcommand\ucb{c} 
\newcommand\trendF{\mathit{f}}

\newcommand\lServed{x}

\newcommand\userSet{\mathbb{U}}
\newcommand\userSetDef{\{\mathbf{\user} \in \{0, 1\}^\nbUsers : \norm{\mathbf{\user}}_1 \leq \maxU \}}
\newcommand\user{u}

\newcommand\bAlignC{\tau}
\newcommand\transRate{\mathit{R}}

\newcommand\pSucc{\mathbb{P}}
\newcommand\transData{d}
\newcommand\reciData{y}

\newcommand\qoe{q}
\newcommand\reward{r}

\newcommand\indicator{\mathbbm{1}}
\newcommand\ind{\indicator_{\textbf{0}}}
\newcommand\resFac{\lambda}
\newcommand\videoBitRate{v_{rate}}
\newcommand\videoRes{v_{res}}
\newcommand\lr{\eta}

\begin{document}

\title{{QoE-Centric Multi-User mmWave Scheduling: \\ A Beam Alignment and Buffer Predictive Approach}
}

{\author{\IEEEauthorblockN{Babak Badnava, Sravan Reddy Chintareddy, Morteza Hashemi}
\IEEEauthorblockA{Department of Electrical Engineering and Computer Science, University of Kansas
}
}}


\maketitle

\begin{abstract}
In this paper, we consider the multi-user scheduling problem in millimeter wave (mmWave) video streaming networks, which comprises a streaming server and several users, each requesting a video stream with a different resolution.
The main objective is to optimize the long-term average quality of experience (QoE) for all users. We tackle this problem by considering the physical layer characteristics of the mmWave network, including the beam alignment overhead due to pencil-beams. To develop an efficient scheduling policy, we leverage the contextual multi-armed bandit (MAB) models to propose a beam alignment overhead and buffer predictive streaming solution, dubbed \algName{}.
The proposed \algName{} algorithm optimally balances the trade-off between the overhead and users' buffer levels, and improves the QoE by reducing the beam alignment overhead for users of higher resolutions. We also provide a theoretical guarantee for our proposed method and prove that it guarantees a sub-liner regret bound. Finally, we examine our proposed framework through extensive simulations. We provide a detailed comparison of the \algName{} against a uniformly random and Round-robin (RR) policies and show that it outperforms both of them in providing a better QoE and fairness. We also analyze the scalability and robustness of the \algName{} algorithm with different network configurations. 
\end{abstract}



\section{Introduction}
\label{sec:Intro}
3GPP broadband wireless standards such as LTE-Advanced and fifth generation (5G) technologies and IEEE wireless standards such as 802.11ad and 802.11ay have enabled high data transfer and data-intensive applications, and are moving towards all-connected small-cell networks. The annual data traffic generated is expected to reach one Zettabyte by $2022$, and is estimated that more than $75\%$ of the world's mobile data traffic will correspond to mobile video streaming~\cite{forecast2019cisco,DAL}. This deluge of data traffic, especially demands for high resolution video streaming on portable mobile devices, will pose significant challenges for the wireless and cellular network providers to meet the quality of experience~(QoE) requirements. In contrast to the quality of service (QoS) that is usually quantified in terms of achieved rate and latency, video streaming QoE depends on several factors such as the resolution of video frames, playback buffer level for each user, number of re-buffering events, and frequency of resolution switches.  

In terms of required infrastructure,  millimeter wave (mmWave) networks are capable of providing multi-Gbps data rates, which makes them suitable to meet the ever-increasing demand for video streaming applications~\cite{rappaport2013millimeter,rangan2014millimeter}. However, unlike omni-directional communications in sub-6 GHz, high data rates in mmWave systems come at the price of large coordination overhead due to highly directional communications needed to compensate for large
channel losses~\cite{han2015large,molisch2017hybrid,heath2016overview}.  Although there are extensive works on providing more efficient beam alignment\footnote{In this paper, beam alignment, collectively, refers to initial beam search, beam tracking, beam refinement, and beam switching.}solutions~\cite{zhou2017throughput, muns2019beam, lee2021user, lee2019beam}, this process still consumes resources that otherwise could have been utilized for high-bit-rate data transfer. 



\begin{figure}[t]
    \centering
    \includegraphics[width=.95\linewidth]{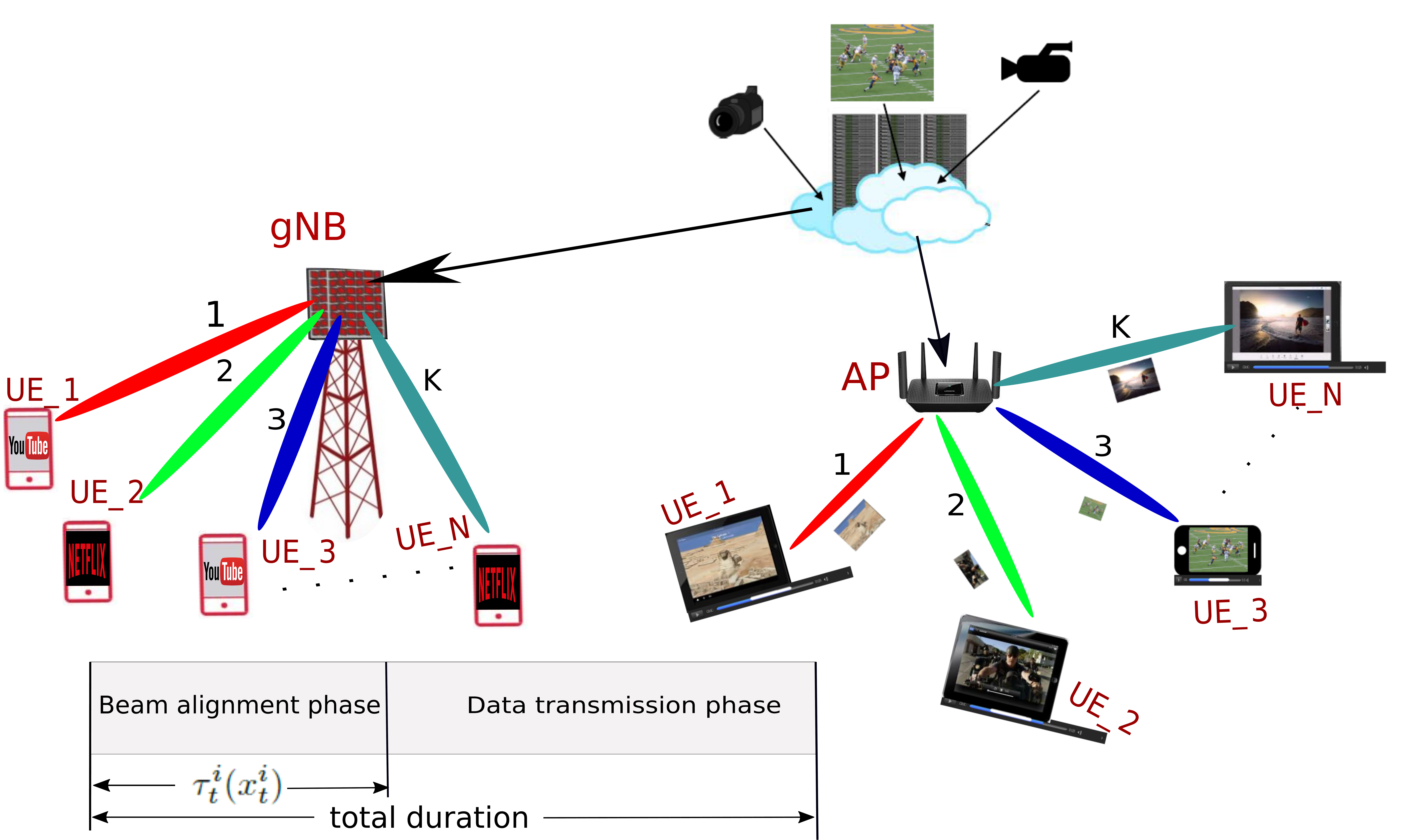}
    \caption{System model depicting mmWave capable base station (gNB) and an Access Point (AP) serving K of N users simultaneously.}
    \label{fig:system_model}
\end{figure}
In mmWave networks, it is true that small wavelengths allow for large antenna arrays to be packed into small chip areas. However, due to power consumption and hardware complexity, the number of RF chains remains limited at the mmWave transceivers, which constrains the number of users that can be served concurrently. A base station, equipped with multiple RF chains ($\maxU$), can serve up to $\maxU$ users at the same time. As such, multi-user management plays a central role to guarantee low-latency and high QoE for all users~\cite{lee2021user, lee2019beam}.
The key point, however, is that due to beam alignment overhead, switching from one mobile user to another one incurs a \emph{switching cost}, as denoted by $\tau(.)$ in Figure~\ref{fig:system_model}. 

Assuming that the QoE is a function of the playback buffer level and resolution of the video frames stored in the buffer, the system needs to balance between users' buffer levels vs. beam alignment overhead to optimize the QoE across all the users in the network. For instance, in the extreme situation, the system could serve only a fixed subset of $\maxU$ users, with the objective of reducing the beam alignment overhead. While this minimizes the risk of zero playback buffer for those $\maxU$ users, the other under-served users would exhaust their playback buffers, which leads to significant QoE degradation. On the other hand, quickly switching between users results in significant beam alignment overheads that is a relatively very slow process vis-à-vis data transfer.


 In this paper, we consider the interplay between beam alignment overhead (i.e., switching cost) and multi-user scheduling in order to enhance the QoE across all users. On one hand, the optimal scheduling should take the switching cost into account, and on the other hand, switching cost is a function of the scheduling algorithm that determines the beam quality. This is in contrast to the classical scheduling problems, where the switching overhead is traditionally assumed to be negligible compared to the service time \cite{hsieh2017delay}.  Indeed, the switching cost and overhead in the mmWave networks becomes critical due to abundant mmWave capacity that would be wasted during the beam alignment phase. 
 
Within this context, and given that beam alignment overhead is a function of the previous schedules, we develop a multi-user scheduling algorithm that works based on predicting the \emph{beam alignment overhead and buffer}. We refer to this algorithm as \algName{} that is built upon the contextual multi-armed bandit (MAB) models to optimally balance the trade-offs between buffer levels and beam alignment overhead. To maximize the average QoE for all users, the streaming server estimates the beam alignment overhead as well as the playback buffer level at each user, and selects $K$ users out of $N$ users at each time slot. There are several studies, such as~\cite{singh2019optimal, adarshtoo, lee2021user,firyaguna2020performance}, that have considered the scheduling task under different scenarios. However, our work aims to integrate the unique characteristics of the mmWave communication (i.e., beam alignment overhead) into a QoE-centric multi-user scheduling framework. In summary, the main contributions of this paper are as follows:
\begin{itemize}
    \item We model the beam alignment overhead of individual users based on the last time we served that particular user, and we propose a dynamic model for the users' buffer level prediction.
    \item Given the playback buffer level, we model the QoE for each user and formulate an optimization problem to improve the long-term average QoE for all the users.
    \item We develop a MAB-based scheduling policy, called \algName{}, which provides a sub-linear regret bound, to solve the defined optimization problem. This algorithm incorporates estimated buffer level of each user and schedule users with the help of a heuristic trend function on the beam alignment overhead.
\end{itemize}

The rest of this paper is organized as follows. In Section \ref{sec:background}, we provide a detailed review of the previous related works. In Section \ref{sec:problem}, we present the system model and formulate a dynamic model of users' playback buffer level. Section \ref{sec:solution} presents the \algName{} algorithm followed by its regret bound analysis. We continue by providing an experimental evaluation of our proposed method in Section \ref{sec:experiments}. 
Finally, Section \ref{sec:conclusion} concludes the paper.

\section{Related works}
\label{sec:background}
 In this section, we review two classes of most related works on user scheduling and QoE optimization.   

 \textbf{Related Works on User Scheduling:} \citeauthor{Jiang2020}~\cite{Jiang2020} proposed a multi-task deep learning scheme for user scheduling and beamforming. They modeled both beamforming and user scheduling as two classification tasks and used a parameter sharing technique to train a multi-task deep neural network. 
\citeauthor{Wu2017ICCC}~\cite{Wu2017ICCC} proposed a two-stage scheduling scheme, based on the inter-user channel correlation and the channel energy, to maximize the sum rate achievable and minimize the overall computational complexity. Although this method maximizes the overall achievable rate, it may provide a poor performance in terms of fairness, since those users with low orthogonality are vulnerable to starvation (e.g., zero buffer size, in our proposed streaming model). 
For uplink scheduling scenarios, \citeauthor{Adan2021Wideband} in ~\cite{Adan2021Wideband} improved their previous work in \cite{Adan2020User} by  developing a distributed quantizer linear coding to cluster the users into different groups such that the number of groups would be as close as to the number of RF chains. Then, they proposed different scheduling schemes based on the users' group. 
The authors in \cite{lee2021user, lee2019beam} develop a policy such that under user mobility, the selected users remain the same unless an abrupt change in the beam direction happens. While this work is closet to our model, it does not consider the QoE, users' buffer level, and impact of frame resolutions for multi-user scheduling in video streaming applications.  

\citeauthor{xu2020joint}~\cite{xu2020joint} proposed a multi-agent reinforcement learning framework for user scheduling and beam selection that minimizes the long-term average network delay while guaranteeing QoS requirements. While they have considered the beam alignment task and QoS satisfaction, there could be discontinuity between the QoS and QoE in multi-user video streaming scenarios.
\citeauthor{Wang2019Stochastic}~\cite{Wang2019Stochastic} proposed a scheduling and resource allocation framework in which they use Lyapunov optimization to maximize a utility function. Any utility measure that is a non-decreasing concave function of transmission rate, such as proportional-fairness and long-term sum-rate, can be utilized for the scheduling. Similar to the previous works, this method does not consider the QoE requirements, and it is specific to analog beamforming with a single RF chain. 



 \textbf{Related Works on QoE Optimization:} On a different note, there is a multitude of prior works for QoE monitoring, approximation, and prediction.  \citeauthor{adarshtoo}~\cite{adarshtoo} proposes a predictive model of QoE based on low-cost QoS measurements like reference signal received power (RSRP) and throughput; They claim that their model accurately predicts re-buffering events and resolution switches more than  $80\%$  of the time. 
The authors in \cite{Yanjun-2017-BWCCA} obtain the QoS parameters, such as packet loss, jitter, and delay using a mmWave NS-3 simulator and devise a non-linear regression method to predict and monitor the QoE of a video streaming service. 
In~\cite{Eswara2020Streaming}, a specific type of recurrent neural networks called LSTM is used to design a QoE predictive model based on features such as playback status, time elapsed since last re-buffering event, and a short term subjective video quality measure. A different work in this category is provided by \citeauthor{Nightingale-2018-I3E-TB} \cite{Nightingale-2018-I3E-TB}. They propose a predictive model of the QoE for an ultra-high-definition (UHD) live video streaming application based on congestion indicators of a 5G network.
However, all these predictive models needs to be incorporated into a decision-making scheduler to prevent the re-buffering, resolution switches, and other undesirable events that degenerate the QoE.


There exist many other works in the literature that approached the QoE requirement satisfaction in different ways. For example, \citeauthor{Li2018QoE}~\cite{Li2018QoE} proposed an optimal cache placement algorithm to maintain a high QoE in an adaptive streaming application; \citeauthor{Tuysuz2020QoE}~\cite{Tuysuz2020QoE} designed a mobility-aware collaborative video streaming client application in which a group of users with different characteristic can stream a high quality video, while maintaining the required QoE; and many other works that are less relevant to our work \cite{Sun2019Delay, Carlsson2017Optimized, Redowan2019Quality}.

Compared with the previous works, our QoE optimization framework at the application layer (i.e., video streaming) takes into account the physical layer characteristics of the mmWave networks and transceiver hardware limitations expressed in terms of the number of RF chains. Indeed, signal directionality is the most distinguished feature for mmWave systems, and thus our multi-user scheduler considers the beam alignment overhead and its impacts on QoE. Overall, our developed solution is tailored for mmWave systems that will be one of the key enablers in the Next-G era. 
\section{System Model and Problem Formulation}



\label{sec:problem}
We consider a mmWave network that consists of $\nbUsers$ users and a single base station (BS) or access point (AP), referred to as the streaming server. At each time slot, the streaming server selects $K$ users out of $N$ users to serve  simultaneously. The $\maxU$ beams generated by the server are used to stream different video frames to each of the $K$ users. Different resolution of each video is available at the server side, and each user may ask for a different resolution based on the channel condition. 

\subsection{Beam Alignment Model and Assumptions}
\begin{figure}[t!]
\centering
    \includegraphics[width=.9\columnwidth, trim= 10mm 10mm 10mm 10mm, clip=true]{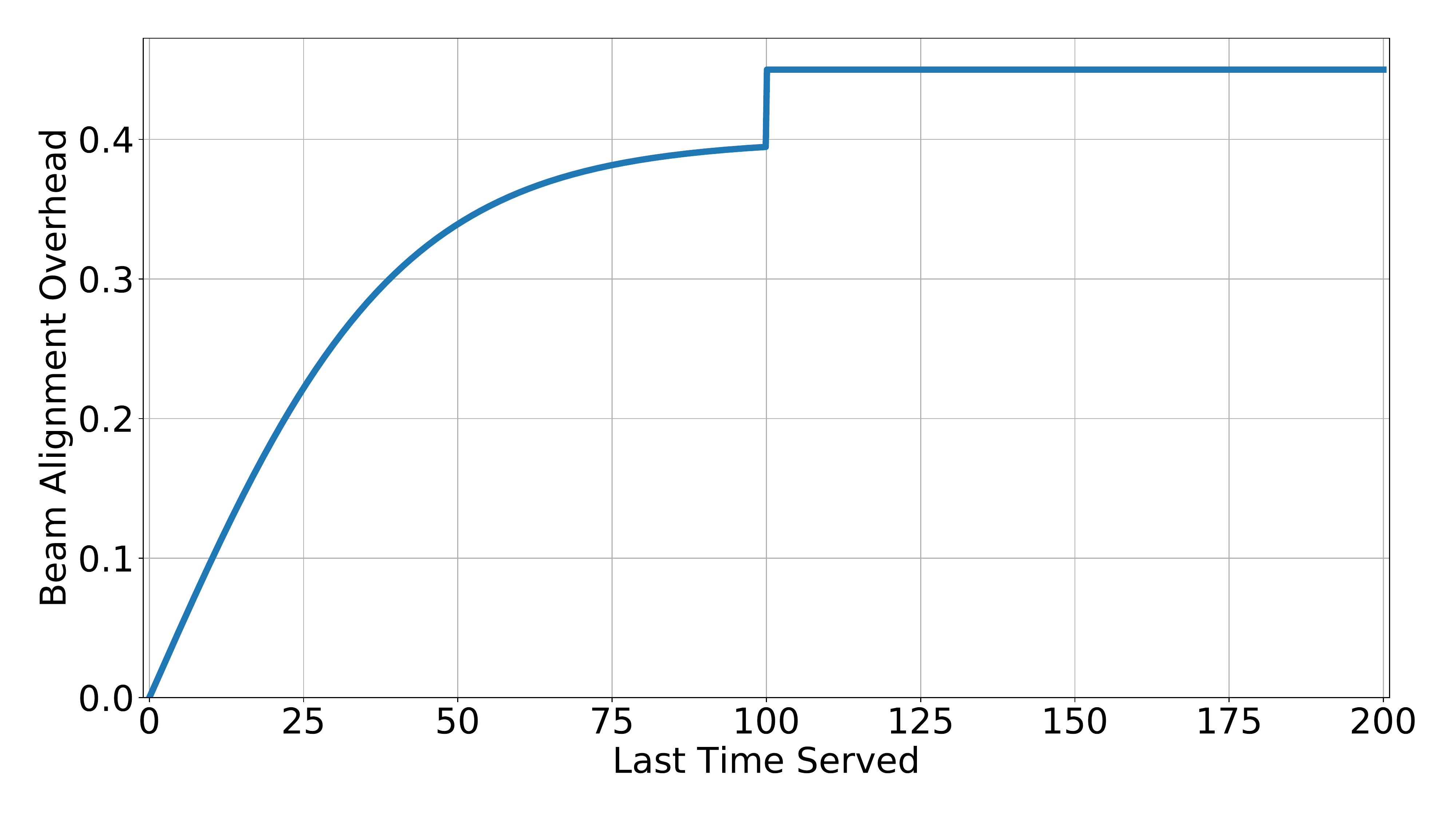}
    \caption{Beam alignment overhead $\tau(.)$ as a function of the time interval between two consecutive schedules. }
    \label{fig:beam-trend}
\end{figure}

As shown in Figure~\ref{fig:system_model}, each time slot is divided into two phases: beam alignment and data transmission. In this paper, we normalize the duration of each time slot to be equal to 1 unit of time. Thus, given that beam alignment takes $\tau(.)$ units,  $1-\tau(.)$ is the amount of time left for the data transfer phase. In the extreme, this overhead can occupy up to $45\%$ of the time slot duration in the cellular networks~\cite{TS38213}. However, in a more general sense, the function $\tau(.)$ can be expressed in terms of  the time interval between two consecutive schedules of a user, i.e., the beam alignment overhead for a user at a specific time depends on how long ago that user was served by the server.  This model captures the ``freshness'' of the beam for the user. 
We can model this characteristic using a \emph{non-decreasing function} of the last time a user has been served. Thus, the beam alignment overhead of user $i$ at time $t$ is denoted by $\bAlignC^i(\lServed_t^i)$, where $\lServed_t^i$ is the amount of time that has been passed since the last schedule of user $i$ at time $t$. 
For instance, Figure \ref{fig:beam-trend} demonstrates such a function for $\tau$. The value of the function $\bAlignC^i(.)$ increases as the $\lServed_t^i$ increases. A full beam search, which consumes $45\%$ of the time slot, is required if it has been passed more than 100 time steps since the last time the user has been served, which is an assumption that can be determined based on the system requirements as suggested in \cite{lee2021user}.






\subsection{Playback Buffer Dynamics}
Each user $i$ has a finite playback buffer of size $\bufferSize^i$ bytes to store video frames. Considering the resolution of the video $\videoRes^i$ that the user is playing and its bit rate $\videoBitRate^i$, the user has at most $\bufferSize^i / \videoBitRate^i$ seconds of video to play. We denote $\bl_t^i$ as the buffer level of user $i$  at time $t$ in seconds.
When the proper beam has been created, the server can start streaming to the user at a specific rate $\transRate$. As such, the amount of data transferred to the user at a time $t$ is obtained as follows: 
\begin{align}
\label{eq:transmitted-data}
\transData_{t}^i = (1 - \bAlignC_t^i (\lServed_t^i)) \transRate / \videoBitRate^i, 
\end{align}
where $\bAlignC_t^i (\lServed_t^i)$ is the beam alignment overhead, and $\transData_{t}^i$ determines the amount of data, in seconds, that the server sends to the user. Due to a blockage and other environmental issues, the user may not receive all the data that has been sent by the server. For the sake of exposition, we assume that the probability of successful reception is given by $\pSucc_t^i$. Therefore, the amount of received data is given by $\reciData_{t}^i = \transData_{t}^i \pSucc_t^i$. The probability value $\pSucc_t^i$ depends on the several factors such as user mobility, blockage, and propagation environment. 

Now that we know how much data a user receives, the dynamics of the playback buffer level is given as follows:
\begin{align}
\label{eq:buffer-dynamics}
\bl_{t+1}^i = \max \{\bl_{t}^i - 1, 0\} + \user_{t}^i\reciData_{t}^i,  
\end{align}
where $\user_{t}^i$ is a binary control variable that determines whether the user $i$ is scheduled at time $t$ or not. Therefore,  $\user_{t}^i\reciData_{t}^i$ determines the amount of seconds of the video that would be successfully transmitted to the user, if it is scheduled. We consider that each time slot is one second, and thus we subtract a second from the previous buffer level of user $i$, and then add the amounts of seconds that the user would receive in case of selection and reception.

\subsection{Quality of Experience}
The QoE for each user depends on the playback buffer level and the resolution of the video frames stored in the buffer. To characterize the QoE, we consider three factors. \textbf{(1)} Any interruption in the streaming is undesirable, and it happens whenever the playback buffer becomes empty. We call this event ``zero-hit''. \textbf{(2)} The QoE increases as the buffer level increases, but it has a diminishing return modeled as a logarithmic function. 
\textbf{(3)} The resolution of the video frames impacts the QoE. For two different users with the same amount of data in their playback buffers, the QoE of the user who plays a higher resolution is higher. We denote the QoE of user $i$ at time $t$ by $\qoe_t^i$, and putting together these factors, the overall QoE can be expressed as: 
\begin{align}
\label{eq:qoe}
\qoe_t^i = (1-\ind(\bl_i^t))\resFac(\videoRes^i) + \alpha \log(1 + \bl_t^i) - \gamma \ind(\bl_i^t). 
\end{align}
The first term is an offset, which only depends on the resolution of the video the user is playing. The second term captures the diminishing return of the playback buffer, and the third term accounts for the zero-hit events that penalizes the QoE by a factor of $\gamma$.

\subsection{QoE-Centric Optimization Problem}
The objective of the server is to maximize the long-term average QoE for all users, given that switching to a new user (a user that was not scheduled in the previous time slot) incurs a beam alignment overhead $\tau(.)$. 
The decision variable is $\mathbf{\user}~\in~\{0, 1\}^{\nbUsers}$, which is a binary vector of size $\nbUsers$. At each time step, only $\maxU$ elements of $\mathbf{\user}$ can be active, and the rest of them are zero. Therefore, we can formulate the following optimization problem:
\begin{equation}
\label{eq:optimization-problem}
   \begin{cases}
   \max\limits_{\mathbf{\user}} \quad & \lim\limits_{\nbTs \to \infty}\frac{1}{\nbTs}\sum\limits_{t=1}^{\nbTs}\sum\limits_{i=1}^{\nbUsers} \qoe_t^i \\
   \textrm{s.t.} \quad & \sum_{i=1}^{\nbUsers} \user^i_t \leq \maxU \,\, \forall_{t= 1..\nbTs }\\
   & \bl_{t+1}^i = \max \{\bl_{t}^i - 1, 0\} + \user_{t}^i\reciData_{t}^i \,\, \forall_{i=1..\nbUsers, t=1..\nbTs} 
   \end{cases}
\end{equation}
The first constraint addresses the hardware limitations in terms of the number of RF chains, and the second constraints captures the playback buffer dynamics. 
The size of the decision space in Eq.~\ref{eq:optimization-problem} scales with the number of users in the network. In the next section, we establish an efficient scheduling framework based on contextual multi-armed bandits. 


\section{Beam and Buffer Predictive Streaming: B2P-Stream}
\label{sec:solution}
In this section, we first provide a brief introduction on multi-armed bandit models. Then, we develop a solution for the problem defined in Eq. \ref{eq:optimization-problem} and present the \algName{} algorithm. Finally, we derive an upper-bound on the regret for the proposed algorithm.

\subsection{Multi-Armed Bandit Models: An Overview}\label{sec:bandit}
A MAB problem is an interactive game between a learner and an environment \cite{bandit-book-csaba}. The game repeats for a finite number of times. In each round of the game, the learner chooses an action (i.e., plays an arm) $\mathbf{\user}$, and receives a reward $\mathbf{\reward}$, that is revealed by the environment. The reward can come from a stochastic distribution or chosen by the environment itself. The learner tries to find an optimal policy using the history of played actions and received rewards. To this end, the Upper Confidence Bound (UCB) method \cite{Li2010ACA, Bouneffouf2016MultiarmedBP} handles the \emph{exploration and exploitation trade-off} by providing an upper bound for the estimation of the expected reward of each arm. The upper bound decreases as the number of reward samples from one arm increases, which means that we are more certain about the estimation of the expected value. There are other classes of MAB algorithms that are specified to different cases, such as the case that the learner can choose more than one arm at a time  \cite{Ontan2017CombinatorialMB, pmlr-v28-chen13a}, called combinatorial bandit problem; or another case where there are some contextual information available \cite{Li2010ACA}.


The performance of bandit models is usually measured in terms of \emph{regret} that quantifies the gap with respect to the optimal solution.  
Let $\expR^i$ be the expected value of the rewards achieved by playing arm $i$ (i.e., $\expR^i = \mathbb{E} (\reward^i)$), and $\expR^{i^*} = \max\limits_i \expR^i$ be the expected value of the reward of the optimal arm. In this case, the immediate regret is defined as $\Delta_i = \expR^{i^*} - \expR^i$, and the accumulated stochastic regret is defined over $\nbTs$ rounds of playing the game, which is given by~\cite{bandit-book-csaba}:
$$R(\nbTs) = T\expR^{i^*} - \mathbb{E}\left[ \sum_{t=1}^{\nbTs}\reward_t\right] =  \sum_{i} \Delta_{i} \mathbb{E}(n_i(T)),$$ 
in which $n_i(T)$ is the number of times that arm $i$ has been played over the time interval $T$.

\subsection{B2P-Stream Policy}
In order to solve the optimization problem defined in Eq. \ref{eq:optimization-problem}, we model this problem as an instance of the \emph{contextual} multi-armed bandit formulation. We designate $\reward_t^i$ as the measurement of the QoE at time $t$ for user $i$, and $\rEst_t^i$ denotes the average of these measurements. In addition, the action set in this model is $\userSet \subseteq \userSetDef$, which tells us that we have a $\nbUsers$ dimensional binary action vector that has at most $\maxU$ active elements.

The contextual bandit model stems from the fact that the scheduler can estimate the playback buffer level at each user as follows: 
\begin{align}
\label{eq:buffer-estimation}
\blEst_{t+1}^i = \max \{\blEst_{t}^i - 1, 0\} + \user_{t}^i\transData_{t}^i.
\end{align}
Note that the state of each user changes over time according to Eq. \ref{eq:buffer-dynamics}, but the scheduler can only estimate the buffer level since there are unknown parameters such as the probability of successful frame reception by the user. 
 This estimated playback buffer level along with the knowledge on beam alignment overhead function $\bAlignC^i(.)$, which is a \emph{non-decreasing function} as a function of the last time served, provide contextual information for the server. For the sake of presentation, we combine these two factors into a single \emph{trend function} $\trendF (\blEst^i)$ that captures the estimated QoE for a user $i$. 
The trend function is then added to the average reward measurements received by the algorithm. In fact, MAB models with trend functions  are finding applications in different domains~\cite{Bouneffouf2016MultiarmedBP}.



\begin{algorithm}[t]
\caption{\algName{}}
\label{alg:bandit}
\textbf{Inputs:} \\
\hspace*{\algorithmicindent}  $\nbUsers$: Number of connected users \\
\hspace*{\algorithmicindent}  $\maxU$: Maximum number of allowed users at each timestamp \\
\hspace*{\algorithmicindent}  $\nbTs$: Total number of timestamps \\
\hspace*{\algorithmicindent}  $\transRate$: The network data transfer rate\\
\hspace*{\algorithmicindent}  $\lr_0$: Initial learning rate \\
\textbf{Algorithm:}
\begin{algorithmic}[1]
\FOR{$t=0$ to $\nbTs$}
	\STATE  $A_t = \emptyset$
	\FOR{$k=1$ to $\maxU$}
		\STATE Select arm $i_k$ = $\argmax_{i \notin A_t} (\rEst_t^i + \ucb_t^i + \trendF^i(\blEst_t) )$
		\STATE $A_t = A_t \cup\{i_k\}$
	\ENDFOR
	\STATE $\mathbf{\user_t} = one\_hot(A_t, \nbUsers)$
	\STATE Perform $\mathbf{\user_t}$ and observe QoE vector $\mathbf{\reward_t}$
	\FOR{$i=1$ to $\nbUsers$}
		\STATE $\transData_{t}^i = (1 - \bAlignC_t^i (\lServed_t^i)) \transRate$
		\STATE $\blEst_{t+1}^i = \max \{\blEst_{t}^i - 1, 0\} + \user_{t}^i\transData_{t}^i$
		\STATE $\rEst_{t+1}^i = \rEst_t^i + \lr_t (\reward_t^i - \rEst_t^i)$
		\IF{$\user_{t}^i == 1$}
		    \STATE $\lServed_{t+1}^i = 0$
		\ELSE
		    \STATE $\lServed_{t+1}^i = \lServed_t^i + 1$
		\ENDIF
	\ENDFOR
	\STATE $\lr_{t+1} = \lr_0 e^{-t / \nbTs}$
\ENDFOR
\end{algorithmic}
\end{algorithm}

The complete process is shown in Algorithm \ref{alg:bandit} in which first we select $\maxU$ users that provide the maximum outcome, and add them to a set $A_t$ (lines 2 to 6). Then, in line 7, we create a $\nbUsers$ dimensional binary vector using $A_t$, and based on this vector, we create $\maxU$ beams and stream to the selected users, and measure the QoE $\mathbf{\reward_t}$. Then using Eq. \ref{eq:buffer-estimation}, we update the playback buffer level estimation in line 11, for all the users. Next, we update the vector $\rEst$ using the new measurements in line 12. Finally, from line 13 to 17, we either set the last time served to zero if we scheduled the user in the current time stamp or increase it if we did not schedule the user. Finally, the learning rate $\lr$ is decreasing exponentially at each iteration.

\subsection{Regret Analysis of \algName{}}\label{sec:regret}
In this section, we provide an analysis of our algorithm and find an upper bound for the regret. In Theorem 1, we show that using a $L_{\trendF}$-Lipschitz trend function, we achieve a sub-linear regret bound for the \algName{} algorithm.

\begin{theorem}
Given an $L_{\trendF}$-Lipschitz trend function, $\bl_{max}$ as the maximum playback buffer level, and an $\alpha > 0$, the regret for \algName{} algorithm is upper-bounded by 
\begin{align*}
R(T) \leq \sum_{i\neq i^*} \frac{2\alpha \log(\nbTs)}{\Delta_i - L_{\trendF} \bl_{max} } + \frac{2\alpha}{\alpha - 1} (\Delta_i + L_{\trendF} \bl_{max}). 
\end{align*}
where $\Delta_i = \expR^{i^*} - \expR^i$.
\end{theorem}

\begin{IEEEproof}
During the learning process, we either underestimate the value of all the sub-optimal actions, event $G_t$, overestimated the value of the optimal action, event $H_t$, or complement of these two events.
\begin{align*}
G_t) \quad & \rEst_{n_i}^i + \trendF (\blEst^i) \leq \expR^i + \trendF (\bl^i) + \ucb^i; \\
H_t) \quad & \rEst_{n_{i^*}}^{i^*} + \trendF (\blEst^{i^*}) \geq \expR^{i^*} + \trendF (\bl^{i^*}) - \ucb^{i^*}, 
\end{align*}
where $\ucb^i = \sqrt{\frac{\alpha \log(t)}{2n_i}}$. Superscript $i$ corresponds to sub-optimal actions and $i^*$ corresponds to the optimal action. We know $G_t$ fails when, 
$
\rEst_{n_i}^i + \trendF (\blEst^i) > \expR^i + \trendF (\blEst^i) + \ucb^i. 
$
By using Hoeffding's inequality, we have
\begin{align*}
\mathbb{P}(G_t^c) & = \mathbb{P}(\rEst_{n_i}^i + \trendF (\blEst^i) - (\expR^i + \trendF (\bl^i)) > \ucb^i) \\
& \leq \exp(-2t(\ucb^i)^2) = \exp(-2t \frac{\alpha \log(t)}{2n_i} ) \\
& \leq \exp(-2t \frac{\alpha \log(t)}{2t} ) = \exp(-\alpha \log(t))  = t^{-\alpha}.
\end{align*}
We achieve the same result for the event $H_t^c$. This result illustrates that as the time passes, the probability of underestimating the optimal action or overestimating all other actions decreases.

\begin{figure*}[!t]
    \centering
     \begin{subfigure}[t]{\textwidth}
         \includegraphics[width=0.32\textwidth, trim= 10mm 10mm 10mm 10mm, clip=true]{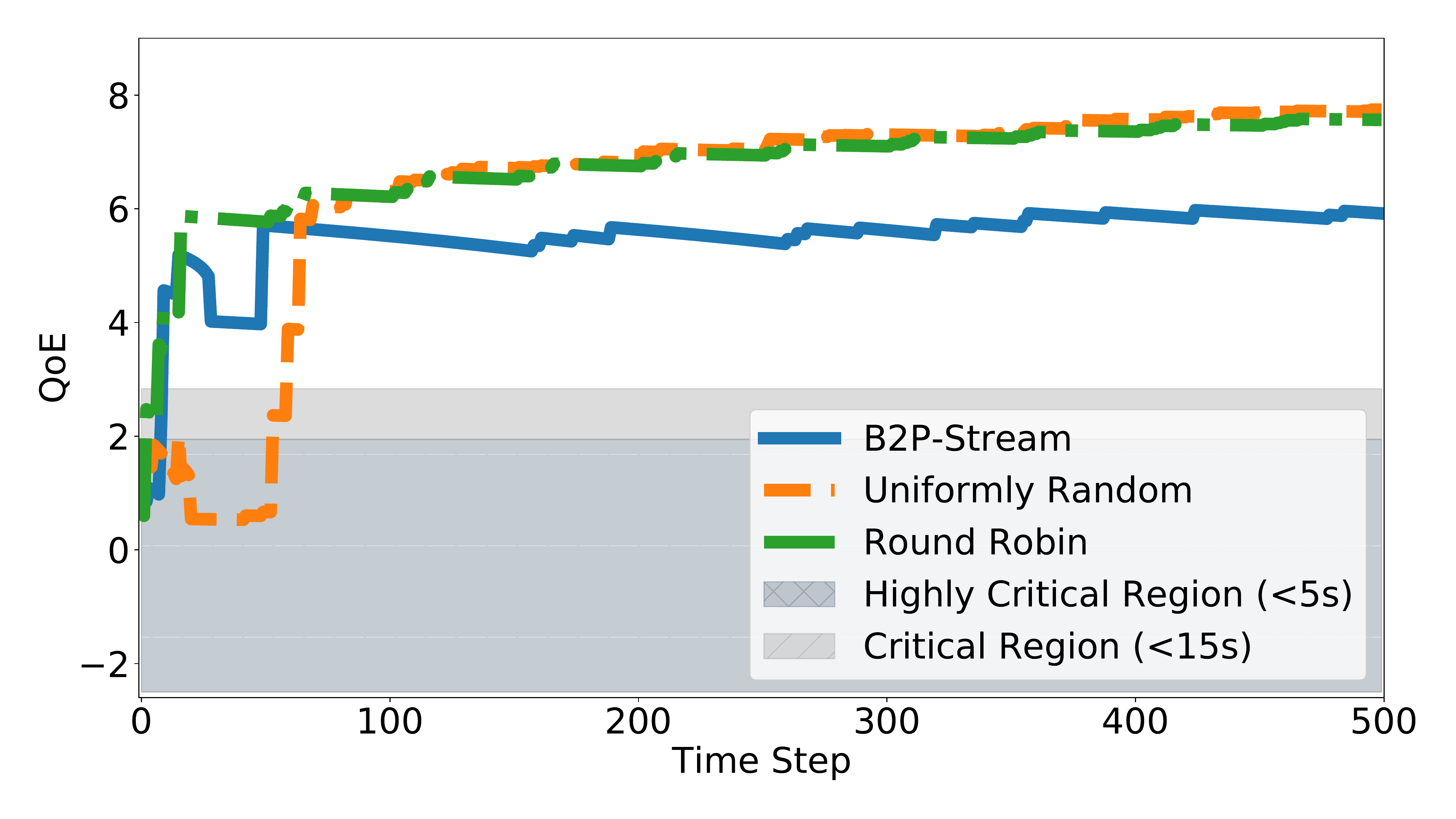}
         \includegraphics[width=0.32\textwidth, trim= 10mm 10mm 10mm 10mm, clip=true]{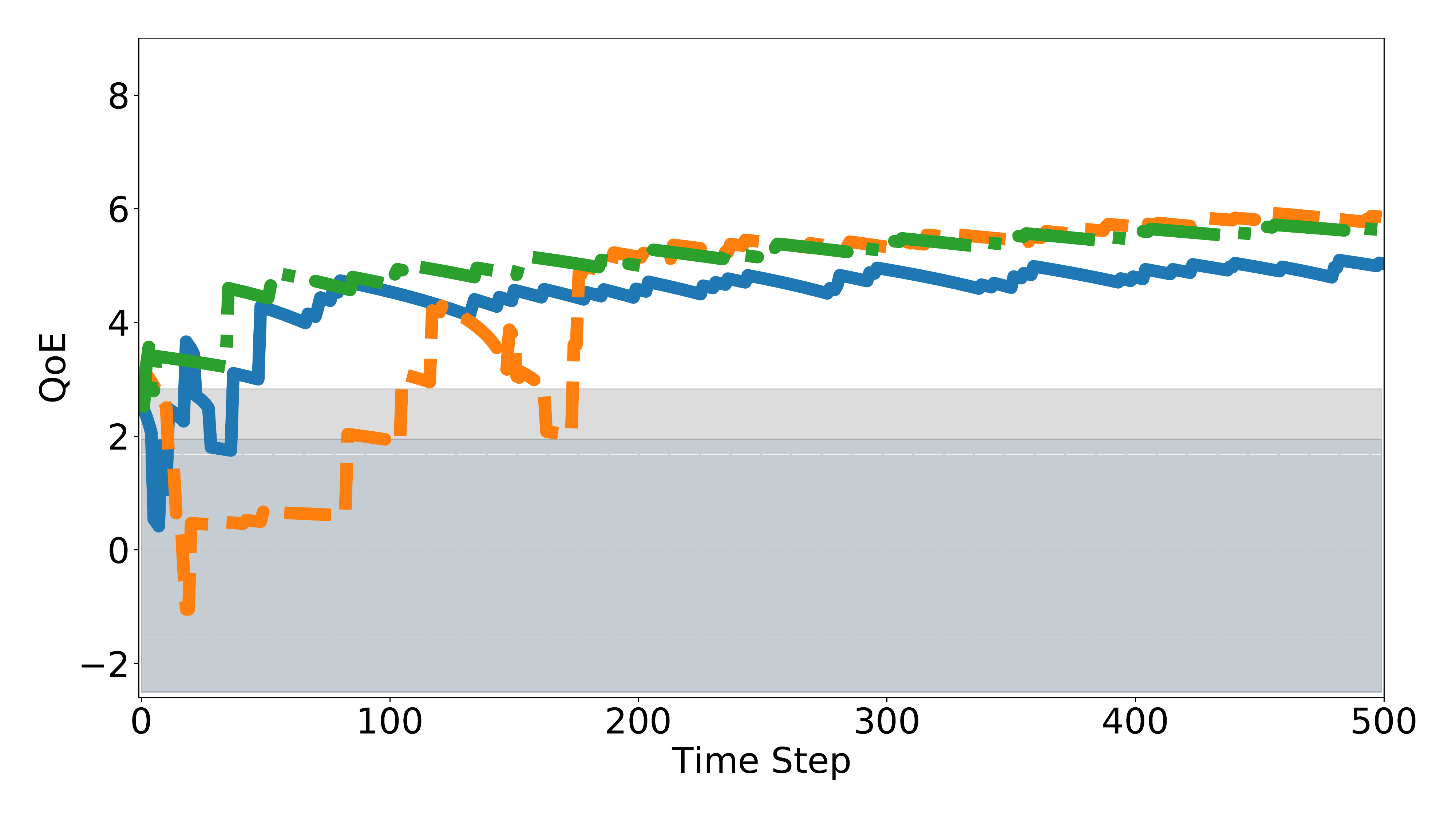}
         \includegraphics[width=0.32\textwidth, trim= 10mm 10mm 10mm 10mm, clip=true]{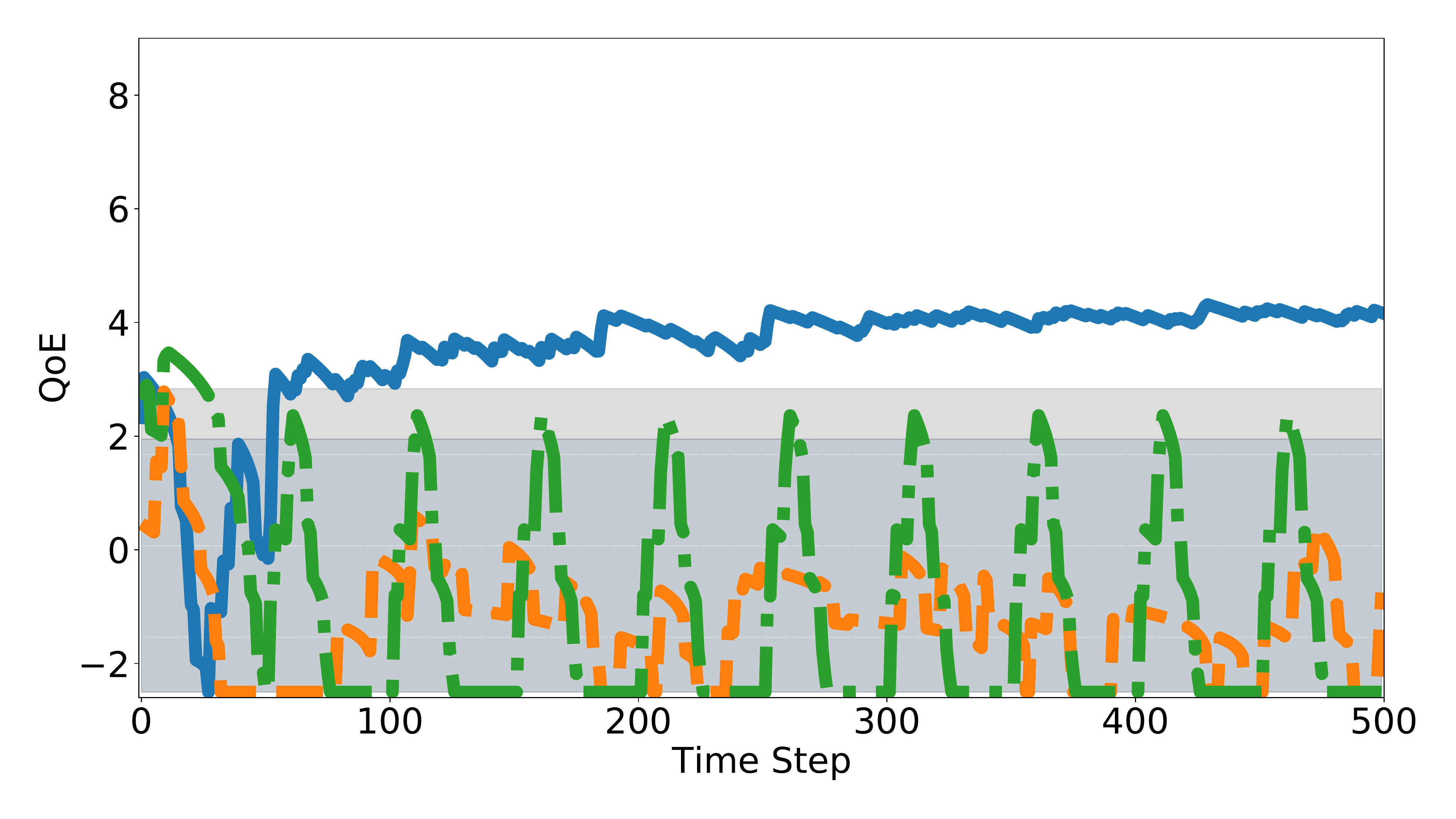}
     \end{subfigure}\hfill
     \begin{subfigure}[m]{\textwidth}
        \centering
        \makebox[20pt]{\raisebox{2pt}{\rotatebox[origin=c]{90}{}}}
         \makebox[0.32\textwidth]{\small{(a) Low Quality (480p)}}
         \makebox[0.31\textwidth]{\small{(b) Medium Quality (1080p)}}
         \makebox[0.31\textwidth]{\small{(c) High Quality (2160p)}}
     \end{subfigure}\hfill
    \caption{The average and standard deviation of measured QoE with 200 users in the network. 
        For lower resolution users, the \algName{} converges faster to be out of the critical regions at the cost of a lower QoE for those users. On the other hand, for users with higher resolutions, the \algName{} provides a faster convergence and better QoE. 
        }
        \label{fig:qoe-diff-algs}
\end{figure*}

Now, let us assume that both $G_t$ and $H_t$ hold. We bound the number of sub-optimal arm pulls. In this case, the sub-optimal arm $i$ is pulled because of insufficient sampling up to this point, which means:
\small{
\begin{align}
\label{eq:insuff-sampling-ineq}
\rEst_{n_i}^i + \trendF (\blEst^i) + \sqrt{\frac{\alpha \log(t)}{2n_i}} \geq \rEst_{n_{i^*}}^{i^*} + \trendF (\bl^{i^*}) + \sqrt{\frac{\alpha \log(t)}{2n_{i^*}}}. 
\end{align}
}
\normalsize
Since $G_t$ and $H_t$ are assumed to be true, by adding $\ucb^i$ and $\ucb^{i^*}$ to both sides of $G_t$ and $H_t$, respectively, we have:
\begin{align}
\label{eq:g_t-plus-epsilon}
\expR^i + \trendF (\bl^i) + 2\sqrt{\frac{\alpha \log(t)}{2n_i}} & \geq \rEst_{n_i}^i + \trendF (\blEst^i) + \sqrt{\frac{\alpha \log(t)}{2n_i}} \\
\label{eq:h_t-plus-epsilon}
\rEst_{n_{i^*}}^{i^*} + \trendF (\blEst^{i^*}) + \sqrt{\frac{\alpha \log(t)}{2n_{i^*}}} & \geq \expR^{i^*} + \trendF (\bl^{i^*})
\end{align}
Now, by chaining equations \ref{eq:insuff-sampling-ineq}, \ref{eq:g_t-plus-epsilon}, \ref{eq:h_t-plus-epsilon}, we have:
\begin{align*}
& \expR^i + \trendF (\bl^i) + 2\sqrt{\frac{\alpha \log(t)}{2n_i}} \geq \expR^{i^*} + \trendF (\bl^{i^*}) \\
&\Rightarrow \sqrt{\frac{\alpha \log(t)}{2n_i}} \geq \frac{\expR^{i^*} + \trendF (\bl^{i^*}) - (\expR^i + \trendF (\bl^i) )}{2} \\
&\Rightarrow \frac{\alpha \log(t)}{2n_i} \geq \frac{(\expR^{i^*} - \expR^i + \trendF (\bl^{i^*}) - \trendF (\bl^i))^2}{4}.
\end{align*}
Let $\Delta_i = \expR^{i^*} - \expR^i$ and $ \delta_i = \trendF (\bl^{i^*}) - \trendF (\bl^i)$, we have:
\begin{align*}
n_i &\leq \frac{2\alpha \log(t)}{(\expR^{i^*} - \expR^i + \trendF (\bl^{i^*}) - \trendF (\bl^i))^2} = \frac{2\alpha \log(t)}{(\Delta_i + \delta_i)^2}.
\end{align*}
Recall that arm $i$ can be pulled when it has been sampled insufficiently (fewer than $\frac{2\alpha \log(t)}{(\Delta_i + \delta_i)^2}$) or either event $G_t$ or $H_t$ fails. Hence, the expected number of times that it has been played is given by:
\small{
\begin{align*}
\mathbb{E}[n_i] &= \sum_{t=1}^{\nbTs} \mathbb{E}[\indicator(I_t = i)]  \leq \frac{2\alpha \log(\nbTs)}{(\Delta_i + \delta_i)^2 } + \sum_{t=1}^{\nbTs} \mathbb{E}[\indicator\{G_t^c \cup H_t^c\}] \\
& \leq \frac{2\alpha \log(\nbTs)}{(\Delta_i + \delta_i)^2 } + \sum_{t=1}^{\nbTs} \Big( \mathbb{E}[\indicator\{G_t^c\}]  + \mathbb{E}[\indicator\{ H_t^c\}] \Big) \\
& \leq \frac{2\alpha \log(\nbTs)}{(\Delta_i + \delta_i)^2 } + \sum_{t=1}^{\nbTs} \Big( t^{-\alpha} + t^{-\alpha} \Big)  \leq \frac{2\alpha \log(\nbTs)}{(\Delta_i + \delta_i)^2 } + \frac{2\alpha}{\alpha - 1}.
\end{align*}
}
\normalsize
Thus, we can bound the regret by
\small{
\begin{align*}
R(T) &= \sum_{i\neq i^*} (\Delta_i + \delta_i)\mathbb{E}[n_i]  \leq \sum_{i\neq i^*} \frac{2\alpha \log(\nbTs)}{\Delta_i + \delta_i } + \frac{2\alpha}{\alpha - 1} (\Delta_i + \delta_i) \\
& \leq \sum_{i\neq i^*} \frac{2\alpha \log(\nbTs)}{\Delta_i + \delta_i } + \frac{2\alpha}{\alpha - 1} (\Delta_i + \delta_i) \\
& \leq \sum_{i\neq i^*} \frac{2\alpha \log(\nbTs)}{\Delta_i - \abs{\delta_i} } + \frac{2\alpha}{\alpha - 1} (\Delta_i + \abs{\delta_i}) \\
& \leq \sum_{i\neq i^*} \frac{2\alpha \log(\nbTs)}{\Delta_i - L_{\trendF} \bl_{max} } + \frac{2\alpha}{\alpha - 1} (\Delta_i + L_{\trendF} \bl_{max}).
\end{align*}
}
\normalsize
Let $\trendF$ be a $L_{\trendF}$-Lipschitz function then we can upper bound $\abs{\delta_i} = \abs{\trendF (\bl^{i^*}) - \trendF (\bl^i)} \leq L_{\trendF} \abs{\bl^{i^*} - \bl^i} \leq L_{\trendF} \bl_{max}$, where $\bl_{max}$ is the maximum level of buffer. 
\end{IEEEproof}

\begin{figure*}[!t]
\centering
    \begin{subfigure}[t]{0.32\textwidth}
         \centering
         \includegraphics[width=\textwidth, trim= 0mm 10mm 10mm 10mm, clip=true]{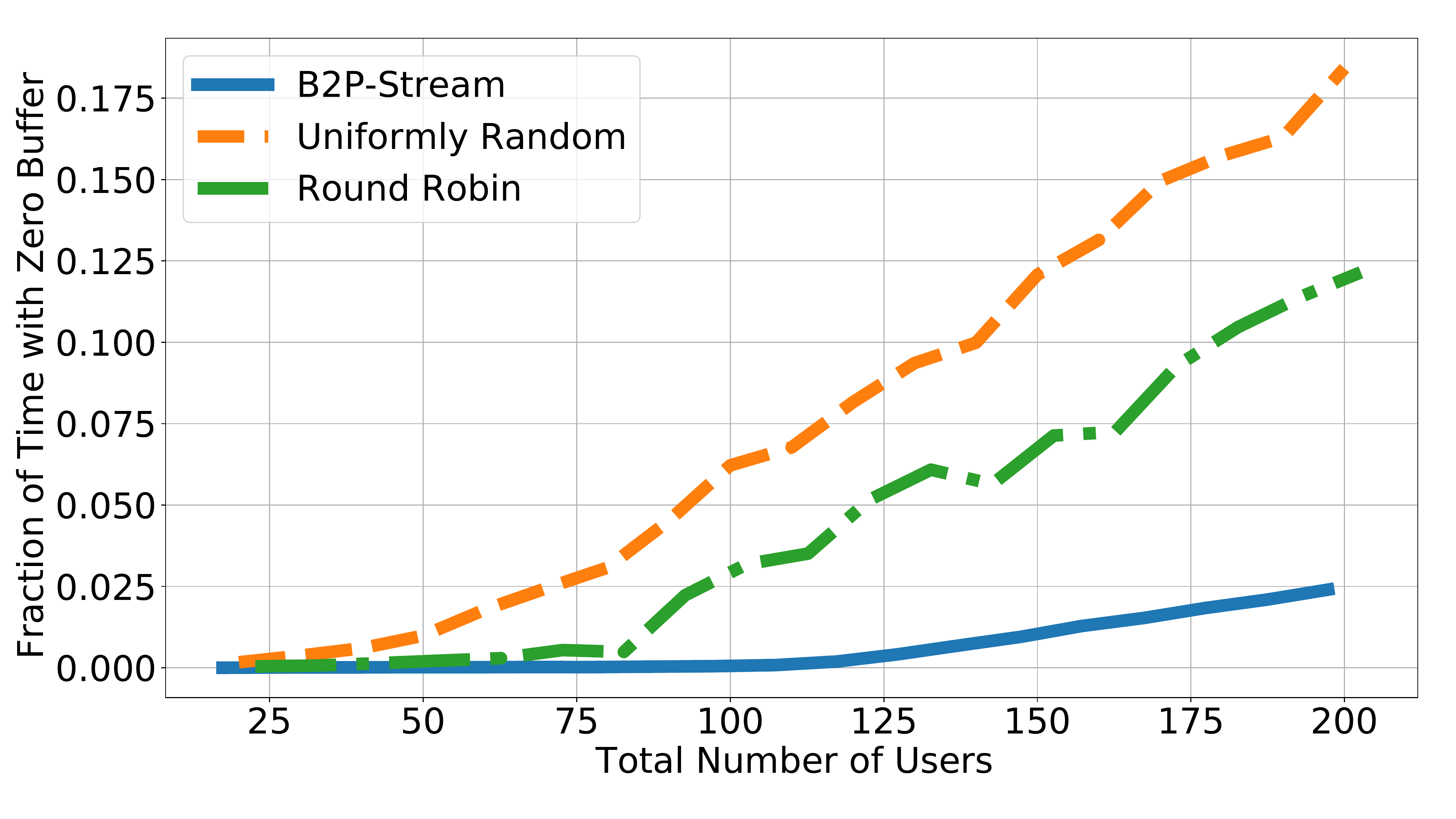}
         \caption{4 RF chains}
         \label{fig:qoe-diff-algs-480}
     \end{subfigure}
     \begin{subfigure}[t]{0.32\textwidth}
         \centering
         \includegraphics[width=\textwidth, trim= 5mm 10mm 10mm 10mm, clip=true]{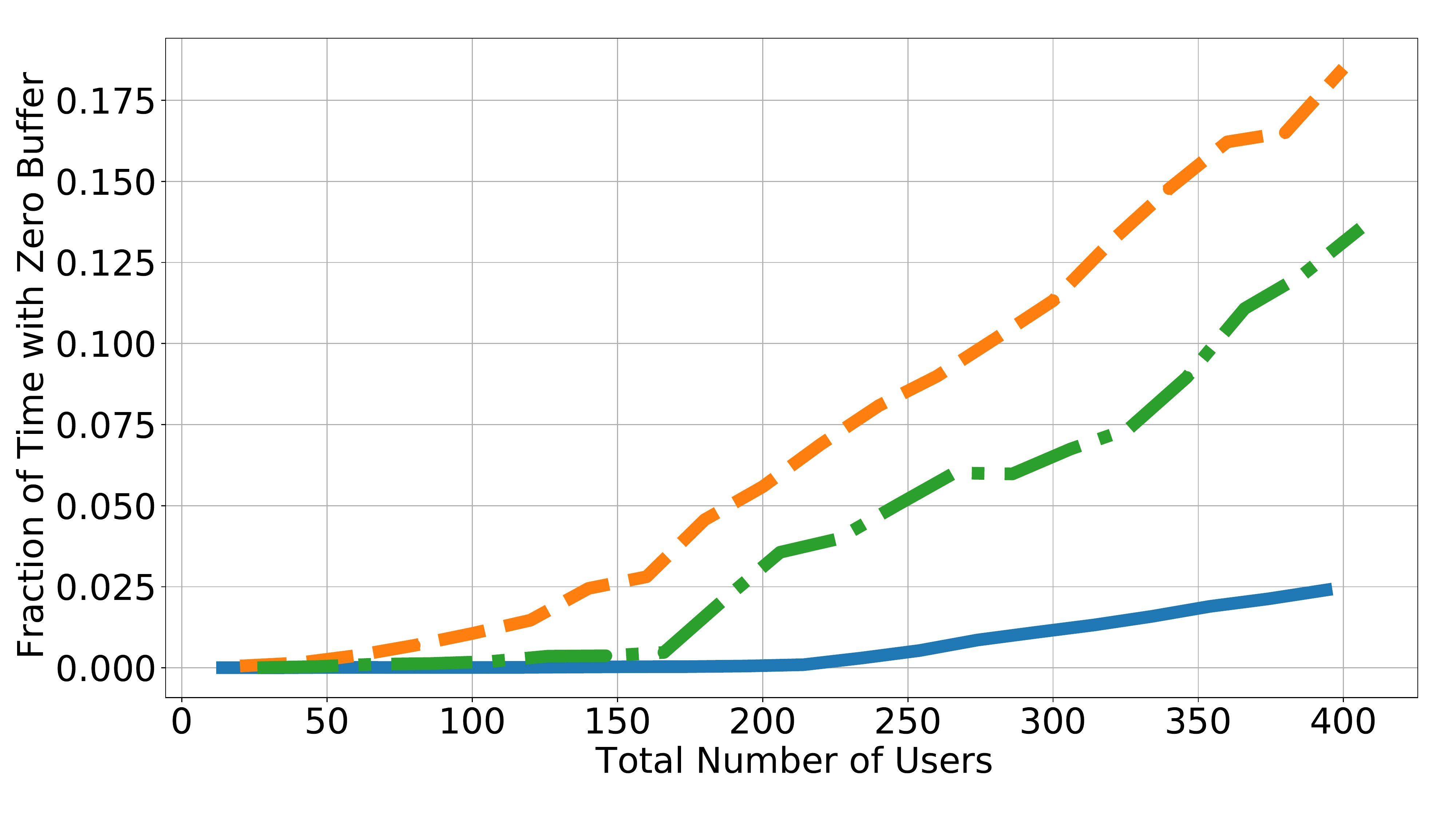}
         \caption{8 RF chains}
         \label{fig:qoe-diff-algs-480}
     \end{subfigure}
     \begin{subfigure}[t]{0.32\textwidth}
         \centering
         \includegraphics[width=\textwidth, trim= 10mm 10mm 10mm 10mm, clip=true]{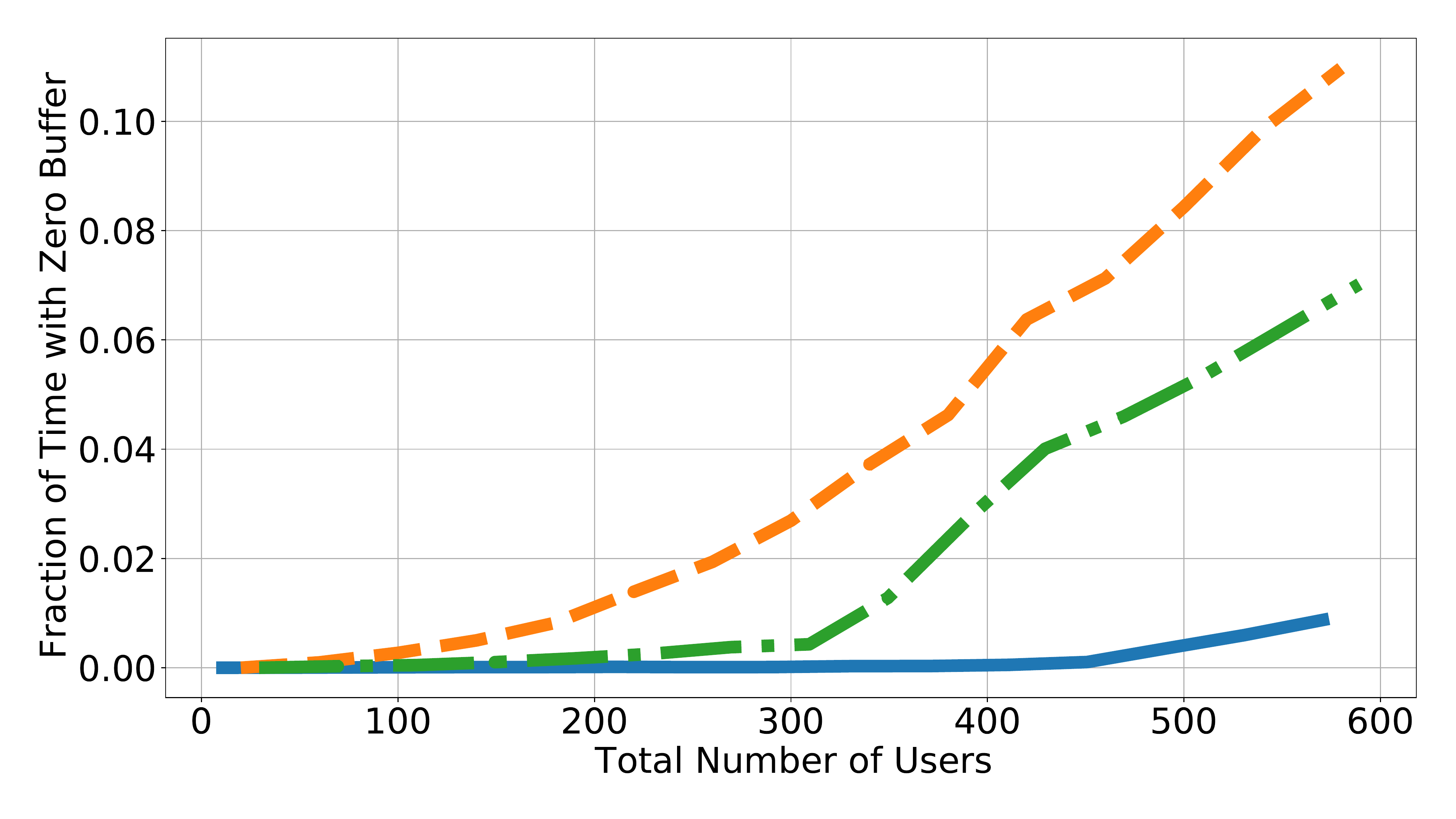}
         \caption{16 RF chains}
         \label{fig:qoe-diff-algs-720}
     \end{subfigure}
     \vspace{-.2cm}
\caption{Fraction of simulation time that users experienced zero-hits per the total number of connected users to the network for \algName{}, Uniform and RR policy. The users scheduled with \algName{} experience far less number of zero-hits.}
\label{fig:zh-algs}
\vspace{-.3cm}
\end{figure*}
\section{Numerical Results}\label{sec:experiments}
In this section, we present the simulation results to demonstrate the efficacy of our algorithm compared with two other baselines.

\subsection{Simulation Setting}\label{sec:sim-setting}
We evaluate the \algName{}'s performance under different conditions to make sure about the robustness of the method. We compare the \algName{} with two different baselines, namely the Uniform and the RR scheduling algorithms. In our simulations, all the users are initialized with zero buffer level. The zero buffer level is considered a highly unsatisfying situation for all the users. Also, the users experience a lower QoE as their playback buffer level approaches zero.
Thus, we define two critical situations to be able to compare the performance of different algorithms.
We call the first critical situation \emph{\textbf{``critical region,''}} that corresponds to the case when a user has less than fifteen seconds of data in the playback buffer. The other one is named \emph{\textbf{``highly critical region,''}} which corresponds to the case when the user has less than five seconds of data in the playback buffer. For instance, 
the dark grey and slate grey, in Figure~\ref{fig:qoe-diff-algs} and Figure~\ref{fig:qoe-diff-N-and-res}, correspond to the critical and highly critical region, respectively.
Since we are initializing all the users with zero buffer levels, it is desirable that the scheduling algorithm avoid these two critical regions. 
\begin{table}[t!]
\center
\begin{tabular}{|c|c|c|}
\hline
\textbf{Resolution} & \textbf{Bit Rate (Mbps)} & \textbf{Portion of Users}\\
\hline
2160p (4K) & 40 & 0.05 \\ 
1440p (2K) &  16 & 0.1\\
1080p &  8 & 0.4\\
720p &  5 &  0.3\\
480p & 2.5 & 0.1\\
360p & 1 & 0.05\\ 
\hline
\end{tabular}
\caption{Video resolution bit rates and the portion of users with a specific resolution.}\label{tb:video-bitrate}
\end{table}

Each experiment has been run 10 times and for 500 time steps. At the beginning of each run, the video resolution requested by each user is selected randomly from the list of resolutions reported in Table \ref{tb:video-bitrate}. The video bit rates and portion of users created for each video quality are also have been set according to Table \ref{tb:video-bitrate}. 
The portion of users with specific resolution is inspired by~\cite{portion4k2018}.

We consider the mean and standard deviation of the observed performance. Since we initialize all the users with a highly critical state, the beginning steps in a simulation are more informative and can reveal more about the underlying events. 

\subsection{QoE Comparison}\label{sec:comparison}
Figure~\ref{fig:qoe-diff-algs} shows the average QoE for $200$ users connected to the network. The BS is equipped with $4$ RF chains and users play videos with different resolutions.
Each of the columns corresponds to a different group of users with different video resolutions: low resolution (480p), medium resolution (1080p), and high resolution (4K), respectively. In this set of results, we ignore the impact of resolution on the QoE (i.e., the first term in Eq. \ref{eq:qoe}) to provide a fair comparison across different resolutions. Also, the dark grey and slate grey, in Figure~\ref{fig:qoe-diff-algs}, correspond to the critical and highly critical region, respectively.
The users scheduled by \algName{} exit the critical regions much earlier than the users scheduled with Uniform or RR policy. Even though only $5\%$ of the users in the simulation are playing a 4K video, both RR and Uniform policies cannot provide a satisfying experience. The situation is the same for users who are playing a 2K video. 

Although the \algName{} provides a better zero-hit statistics compared to two other baselines, there is some cost needs to be paid. This cost is lower QoE for users of lower resolution. From the results, we note that the \algName{} algorithm maintains a lower QoE for users of lower resolution to compensate for the users of higher resolutions, as they need to be served more often because of the higher bit rate requirements. This does not mean that users with lower video quality would experience a significantly lower QoE because the QoE has a diminishing effect and there is not much of QoE difference as long as they are out of critical regions.

\subsection{Zero-hit Performance}
Next, we compare the average fraction of simulation time that each user experiences a zero-hit when the streaming server is equipped with $4$, $8$, or $16$ RF chains. The number of RF chains determines the maximum number of users that can be scheduled simultaneously at each time slot. Figure~\ref{fig:zh-algs} compares the performance of \algName{} with respect to other baselines, as the number of RF chains and total number of users increase.  From the results, \algName{} achieves a smaller zero-hit compared with the RR and Uniform scheduler, as the number of RF chains is equal to $4$, $8$, or $16$. 



\subsection{Scalability Analysis} \label{sec:parameter-evaluation}

\begin{figure}[t!]
    \centering
    \includegraphics[width=\columnwidth, trim = 22mm 0mm 22mm 10mm, clip=true]{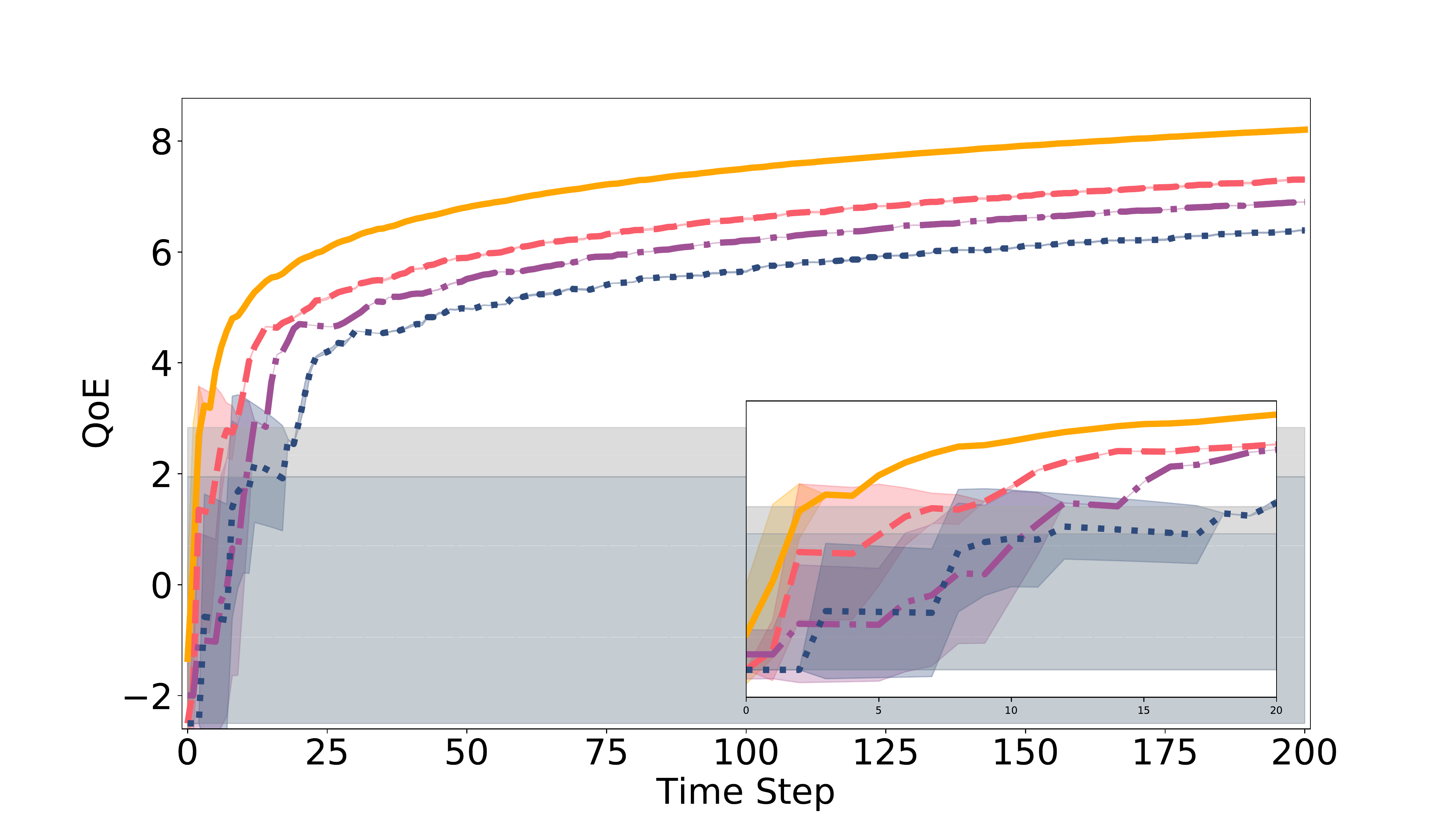}
    \includegraphics[width=\columnwidth, trim = 0mm 10mm 0mm 10mm, clip=true]{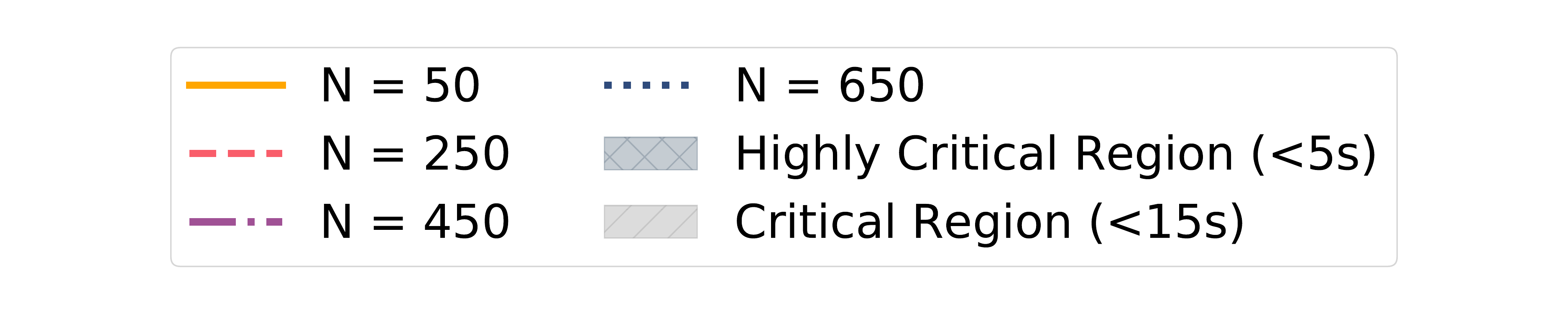}
     \caption{The average and standard deviation of the measured QoE achieved by \algName{}, as the total number of connected users ($N$) is changing. 
     }
     \label{fig:qoe-diff-N-and-res}
\end{figure}\

Figure~\ref{fig:qoe-diff-N-and-res} reveals more about the scalability and reliability of \algName{}. This figure reports the average and standard deviation of the measured QoE, but it is specific to the users who are streaming a 4K video, since they are more prone to unsatisfying QoE. Figure~\ref{fig:qoe-diff-N-and-res} demonstrates how the QoE behaves for these users as the total number of connected users increases. From the results, we note that even with $N = 650$ connected users and  $K = 4$ RF chains, it takes less than $25$ time steps for \algName{} to push all the high quality users out of the critical regions and provide an improved QoE performance. 

     


\subsection{Beam Alignment Overhead Results}
\begin{figure}[t!]
    \centering
    \vspace{.3cm}
    \includegraphics[width=\columnwidth, trim= 5mm 10mm 10mm 5mm, clip=true]{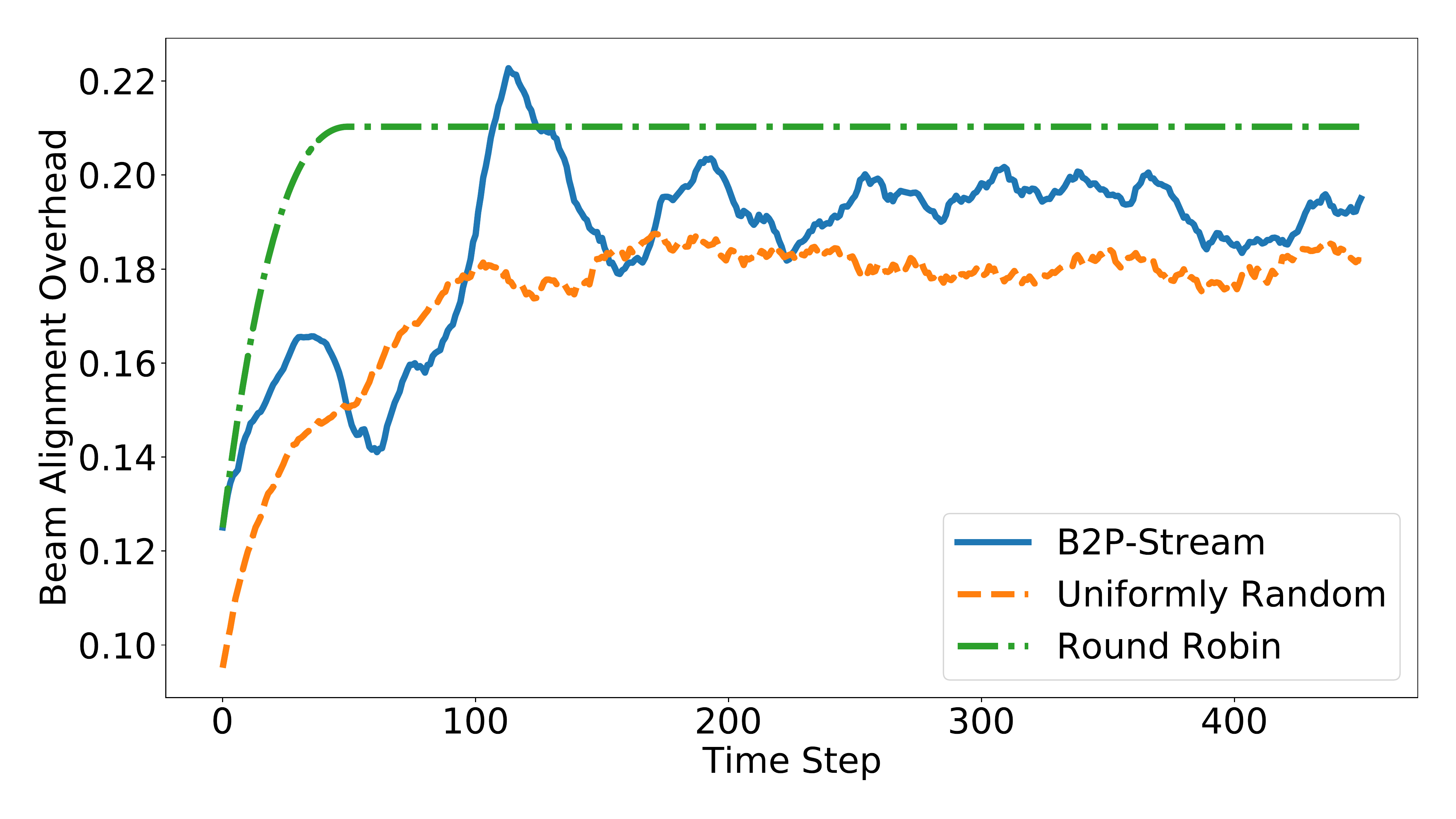}
    \caption{Average beam alignment overhead when $4$ users are allowed to be scheduled at a time. The shaded area shows the standard deviation. 
    }
    \label{fig:beam-overhead}
\end{figure}

Figure~\ref{fig:beam-overhead} shows the average of beam alignment overhead of \algName{} compared to other algorithms. 
A moving average with windows size of 50 has been applied to these curves to make them smoother and more  comparable. The RR algorithm maintains a fixed beam alignment overhead due to its deterministic nature. On the other hand, the Uniform policy maintains a lower beam alignment overhead compared to \algName{} and RR, but as it is depicted in Figure~\ref{fig:qoe-diff-algs}, it fails to satisfy the fairness criterion. Note that although the \algName{} has a slightly larger beam alignment overhead compared to Uniform policy, it cleverly exploits the time resources to provide a fair QoE for all the users. 

\subsection{Intuitions Behind \algName{}}
\begin{figure}[t!]
    \center
    \includegraphics[width=\columnwidth, trim= 10mm 10mm 10mm 10mm, clip=true]{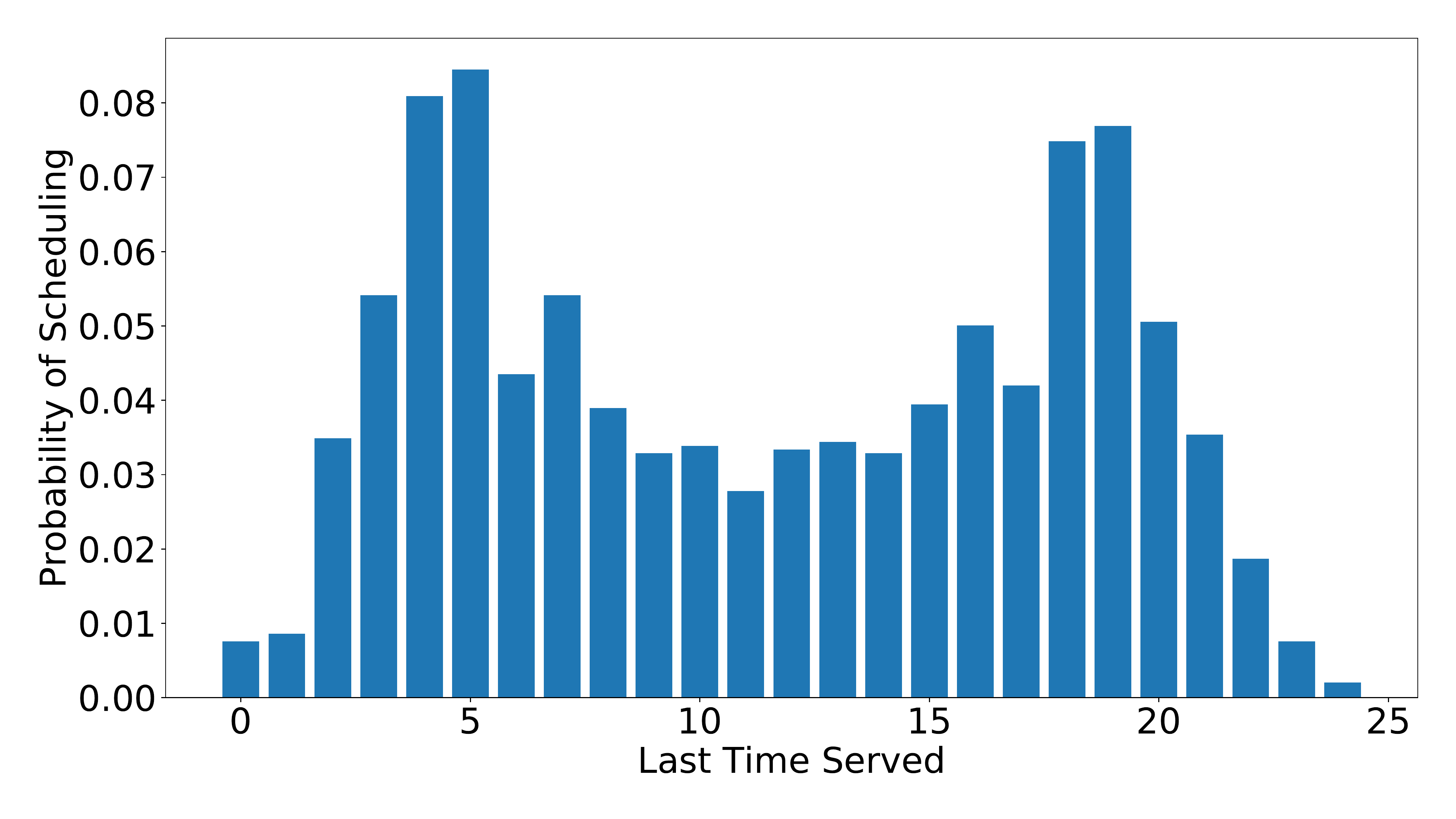}
    \caption{Empirical distribution of the scheduled users with respect to the last time they have been served. The first peak corresponds to the users with a lower beam alignment cost, and the second peak corresponds to the users with an exhausted buffer level.}
    \label{fig:lServed-dist}
\end{figure}

To balance the trade-off between beam alignment overhead and playback buffer levels and optimize the QoE metric, we can intuitively distinguish two groups of users. The first group corresponds to those users who were served recently, thus the beam alignment overhead would be small for them (small $\bAlignC^i_t$), and we can stream more data to this group. The second group are those users that their buffer levels are approaching zero, meaning that it has been a long time since the last time we served them (large $\lServed^i_t$). Figure~\ref{fig:lServed-dist} demonstrates the empirical distribution of scheduled users with respect to the last time those users were served, i.e., $\lServed^i_t$. From the results, we can identify these two groups that correspond to the two peaks in Figure \ref{fig:lServed-dist}, respectively.

\begin{figure}[!t]
    \center
    \includegraphics[width=\columnwidth, trim= 10mm 10mm 10mm 10mm, clip=true]{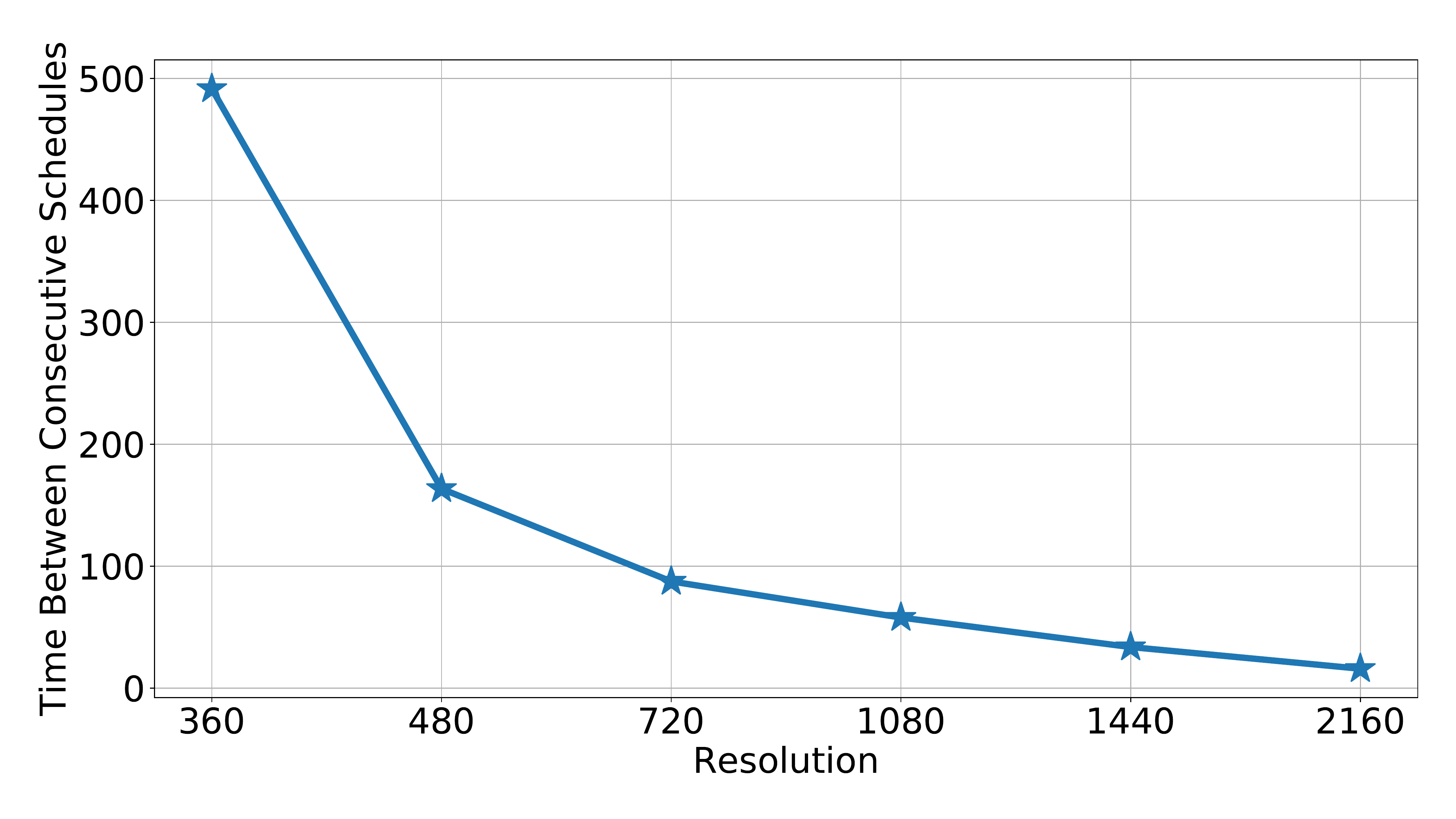}
    \caption{Time interval between consecutive schedules of users of different video resolution. The interval decreases as the resolution increases, means we choose users with higher resolution more frequently.}
    \label{fig:scheduling-frequency}
\end{figure}

Figure~\ref{fig:scheduling-frequency} illustrates the time interval between two consecutive schedules of different groups of users in the simulation. The \algName{} may allocate more resources to the users with higher resolution, which means they would be scheduled more frequently, due to the fact that the video bit rate for them is higher, and they would need more data to play a video for a specific period of time compared with users with lower video quality. We can again see that even though only $5\%$ of the users play a 4K video, they would be scheduled almost every $17$ time steps on average, because their QoE requirement are higher than other users. On the other hand, users with a medium video quality of 1080p, which applies to 40\% of the users, would be scheduled every $62$ time steps.
\section{Conclusion}\label{sec:conclusion}
In this paper, we considered the problem of multi-user mmWave scheduling ($\maxU$ users out of $\nbUsers$) who are streaming videos with different resolutions. The overall objective is to optimize the QoE across all the users. Leveraging the contextual MAB models, we developed a QoE-centric scheduling policy that considers the physical layer characteristics of the mmWave networks. The proposed \algName{} algorithm is able to optimally balance the trade-off between the beam alignment overhead and the users' playback buffer level. 
 In particular, \algName{} uses an estimated buffer level as an input for a \emph{trend function} that biases the scheduling policy towards those users with exhausted buffer levels. We provided theoretical analysis to prove that the \algName{} guarantees a sub-linear regret bound, and through simulations, we showed that the proposed algorithm outperforms both Uniform and RR policies. Overall, mmWave networks are considered as one of the key enablers for data-intensive applications such as high quality video streaming. As such, developing efficient and reliable multi-user management algorithms that guarantee high QoE for all the users, is of utmost importance to enable ubiquitous  mmWave technologies.  




\Urlmuskip=0mu plus 1mu\relax
\bibliographystyle{IEEEtranN}
\bibliography{references}
\end{document}